\documentclass[fleqn,usenatbib]{mnras}

\usepackage{newtxtext,newtxmath}

\usepackage[T1]{fontenc}

\DeclareRobustCommand{\VAN}[3]{#2}
\let\VANthebibliography\thebibliography
\def\thebibliography{\DeclareRobustCommand{\VAN}[3]{##3}\VANthebibliography}

\usepackage[addedmarkup=bf]{changes}

\usepackage{gensymb}
\usepackage{graphicx}	
\usepackage{amsmath}	

\title[Her X-1 triaxial free precession]{Evidence for neutron star triaxial free precession in Her X-1 from \textit{Fermi}/GBM pulse period measurements}

\author[D. Kolesnikov et al.]{
Dmitry Kolesnikov,$^{1}$\thanks{E-mail: kolesnikovkda@gmail.com}
Nikolai Shakura$^{1}$
and Konstantin Postnov$^{1, 2}$
\\
$^{1}$Moscow State University, Sternberg Astronomical Institute, Universitetskij pr. 13, 119234 Moscow, Russia\\
$^{2}$Kazan Federal University, Kremlyovskaya 18, 420008 Kazan, Russia\\
}

\date{Accepted XXX. Received YYY; in original form ZZZ}

\pubyear{2021}

\begin{document}
\label{firstpage}
\pagerange{\pageref{firstpage}--\pageref{lastpage}}
\maketitle

\begin{abstract}
Her X-1/HZ Her is one of the best studied accreting X-ray pulsars. In addition to the pulsating and orbital periods, the X-ray and optical light curves of the source exhibit an almost periodic 35-day variability caused by a precessing accretion disk. The nature of the observed long-term stability of the 35-day cycle has been debatable. The X-ray pulse frequency of Her X-1 measured by the  \textit{Fermi}/GBM  demonstrates periodical variations with X-ray flux at the Main-on state of the source. We explain the observed periodic sub-microsecond pulse frequency changes by the free precession of a triaxial neutron star with parameters previously inferred from an independent analysis of the X-ray pulse evolution over the 35-day cycle. In the \textit{Fermi}/GBM data, we identified several time intervals with a duration of half a year or longer where the neutron star precession period describing the pulse frequency variations does not change. We found that the NS precession period varies within one per cent in different intervals. Such variations in the free precession period on a year time scale can be explained by $\lesssim 1\%$ changes in the fractional difference between the triaxial neutron star's moments of inertia due to the accreted mass readjustment or variable internal coupling of the neutron star crust with the core.
\end{abstract}

\begin{keywords}
X-rays: binaries -- X-rays: individual: Her X-1 -- stars: neutron
\end{keywords}



\section{Introduction}
Her X-1 is an accreting X-ray pulsar with a pulse period of $P^*=1.24$ s around the optical star HZ Her with an orbital period of 1.7 days \citep{1972ApJ...174L.143T, 1972IBVS..720....1C}. The binary system is viewed almost edge-on. This causes different eclipsing features, including periodic orbital eclipses by the optical star and X-ray dips due to gas streams shielding the line of sight \citep[ e.g.,][]{1999A&A...348..917S}.
The source also demonstrates a long-term 35-day X-ray flux modulation  \citep{1973ApJ...184..227G}. It consists of an X-ray bright Main-on state lasting about seven binary orbital periods, followed by a first low state with an almost zero flux (about four orbits), a Short-on state less prominent than the Main-on  (about four orbits), and a second low-on state (about four orbits), see \cite{1998MNRAS.300..992S, 2020ApJ...902..146L} for more detail. 

The nature of the 35-day modulation has been debatable. One of the first explanation involved a
freely precessing neutron star (NS)  \cite{1972Natur.239..325B, 1973SvA....17..295N}. For the observed 35-day period to be the NS free precession period $P_\mathrm{pr}$, an axially symmetric NS should maintain a tiny ellipticity of the order of $\Delta I/I \sim P^*/P_\mathrm{pr} \sim 10^{-6}$ (here $\Delta I$ is the difference in the NS's moments of inertia). In the case of a single NS, the unavoidable internal dissipation would tend to secularly align the spin and precession axes. This argument has been considered disfavoring the NS free precession as the reason for the long-term periodicity in pulsars (e.g., \cite{1977ApJ...214..251S}). The precession of an accretion disk around NS provides another explanation to the 35-day cycle  \citep[e.g.,][and subsequent papers]{1973NPhS..246...87K,  1974ApJ...187..575R, 1975ApJ...201L..61P}. Presently, a rich phenomenology, both in the X-ray and optical, supports the presence of a tilted, retrograde, precessing accretion disk in Her X-1 \citep[e.g.,][]{1973ApJ...186..617B, 2003MNRAS.342..446L, 2006AstL...32..804K,2021ApJ...909..186B}. In the middle of the Main-on and Short-on states, the disk is 
maximum open to the observer's view, while during the low states, the outer parts of the tilted disk block the  X-ray source.

Extensive X-ray observations of Her X-1 demonstrate that there can occur long (with a duration of up to 1.5 years) anomalous low states of the X-ray source during which the X-ray flux is completely extinguished but the X-ray irradiation of the optical star HZ Her persists \citep{1985Natur.313..119P, 1994ApJ...436L...9V, 2000ApJ...543..351C, 2004ATel..307....1B, 2004ApJ...606L.135S}. These anomalous low states are likely  due to vanishing the disk tilt to the orbital plane. As long as the disk tilt is close to zero, the X-ray source remains blocked from the observer's view by the disk's outer parts.
An analysis of archive optical observations of HZ Her using photo plates showed that in the past there were periods when the X-ray irradiation effect was absent altogether \citep{1973ApJ...182L.109J, 1976BAICz..27..325H}. This means that sometimes in Her X-1/HZ Her binary system, the accretion onto the neutron star can cease completely \citep{1978SvAL....4...43B}.
The cessation of accretion could occur, for example, because of a sudden jump
in the NS magnetic field, which sometimes are observed in Her X-1 \citep{2019A&A...622A..61S}, 
or a decrease in the mass inflow from the optical star, which can turn-off accretion due to the propeller effect.

The fact that the 35-day cycle re-appears in phase with the average 35-day ephemeris after the end of anomalous low states and the stable periodic behavior of X-ray pulse profiles \cite{2013A&A...550A.110S} requires a `stable clock' mechanism operating in Her X-1/HZ Her \citep{2009A&A...494.1025S}, which may be the NS free precession. Indeed, a model of two-axial NS free precession can reproduce the observed regular  X-ray pulse profile changes with the 35-day phase \citep{2013MNRAS.435.1147P}. This model involves a complex non-dipole magnetic field structure near the surface of accreting NS in Her X-1 and pencil-beam local emitting diagram. The non-dipole surface fields includes an additional quadrupole component producing ring-like structures around the NS magnetic poles \citep{1991SvAL...17..339S}. This additional field doesn't distort the NS's form which is assumed to be shaped by a much stronger internal magnetic field $\sim 10^{14}$ G \citep{2009MNRAS.397..763B}. The model also can explain the complicated optical variability of HZ Her over the 35-day cycle, which is primarily shaped by the irradiation effect of the optical star's atmosphere by the X-ray emission from NS \citep{2020MNRAS.499.1747K}.
A triaxial NS precession in Her X-1 was proposed earlier by us \citep{1998A&A...331L..37S} to explain an anomalously narrow 35-day cycle of Her X-1 observed by HEAO-1. Presently, there is a growing empirical evidence that  NS free precession could be responsible for different long-term periodicities in single magnetized NS, such as magnetars and fast radio bursts (FRBs) \citep[see, e.g.][]{2020ApJ...895L..30L, 2020ApJ...892L..15Z, 2021arXiv210712874C, 2021arXiv210712911W, 2021arXiv210911150M}.

A precessing, pulsating NS should exhibit regular pulse period (or frequency) variations with a  fractional amplitude change of $\Delta P^*/P^*\approx P^*/P_\mathrm{pr}\sim 10^{-6}$  \citep[e.g,][]{1970Natur.225..838R, 1986ApJ...300L..63T, 1989A&A...221L...7B, 1993A&A...267L..43B, 1995pns..book...55S}. This tiny pulse frequency variations of Her X-1 can be searched for by the continuous monitoring of X-ray sources. 

In this paper, we show that the periodic sub-microsecond pulse period variability observed in Her X-1 at the 35-day cycle maxima (the Main-on state) by \textit{Fermi}/GBM (Gamma-ray Burst Monitor) \citep{2009ApJ...702..791M} can be explained by the motion of X-ray emitting region on the NS surface during the free precession of a triaxial NS. A preliminary analysis of the \textit{Fermi}/GBM data for the two-axial NS free precession was reported in \cite{2021ARep...65.1039S}.

\section{\textit{Fermi}/GBM X-ray pulsar Her~X-1 frequency measurements}\label{sec:frequency-measurements}
\textit{Fermi}/GBM X-ray pulsar Her X-1 frequency measurements are publicly available\footnote{\url{https://gammaray.nsstc.nasa.gov/gbm/science/pulsars/lightcurves/herx1.html}} and updated on daily basis. The measured frequency $\nu(t)$ of Her X-1 can be represented as a sum of non-periodic long-term frequency variability $\nu_0(t)$ and periodic 35-day frequency variability $\delta \nu(t)$:
\begin{equation}
\label{e:nu}
    \nu(t) = \nu_0(t) + \delta\nu(t)
\end{equation}
or, equivalently, 
in terms of the angular frequency:
\begin{equation}
    \Omega(t) = \Omega_0(t) + \delta\Omega(t)\,.
\end{equation}
In accreting pulsars like Her X-1, the long-term pulsar frequency trend $\Omega_{0}(t)$ can be due to changing accretion torques, see Fig. \ref{f:pulsar-frequency}. 

Assuming that the pulsating flux is emitted near the north magnetic pole $N$ of a rotating solid body, the 35-day periodic variations $\delta\Omega(t)$ are defined by the rate of change of the angle $\Phi$ of the spherical triangle $I_3 \Omega N$, see Fig. \ref{f:neutron_star}:
\begin{equation}
    \delta\Omega(t) = \frac{d\Phi(t)}{dt}\,.
\end{equation}
Here we show that the periodic change of the angle $\Phi$ with parameters as in Her X-1 can be explained by a freely precessing NS. We start with considering a two-axial NS precession, which can be treated analytically, and continue with a more general case of triaxial NS free precession.

\section{Free precession of the neutron star}\label{sec:free-precession}

\subsection{Precession of an axially symmetric NS}

\begin{figure}
   \centering
   \includegraphics[width=0.8\hsize]{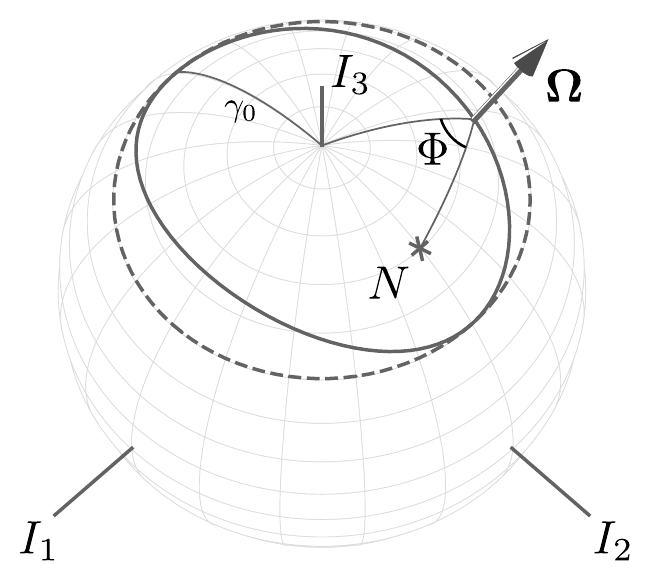}
      \caption{A schematic view of free precession of a triaxial neutron star. The surface coordinates (light grey lines) are related to NS inertial axes $I_1, I_2, I_3$. The path of the NS angular momentum vector (the spin axis) on the NS surface $\boldsymbol{\Omega}$ during the  triaxial free precession is shown by the solid line. The path of  $\boldsymbol{\Omega}$ during a two-axial free precession is shown by the dashed line. The asterisk marks the north magnetic pole $N$. $\Phi$ is the phase angle of the north magnetic pole $N$.}
         \label{f:neutron_star}
   \end{figure}

\begin{figure*}
    \centering
    \includegraphics[width=\hsize]{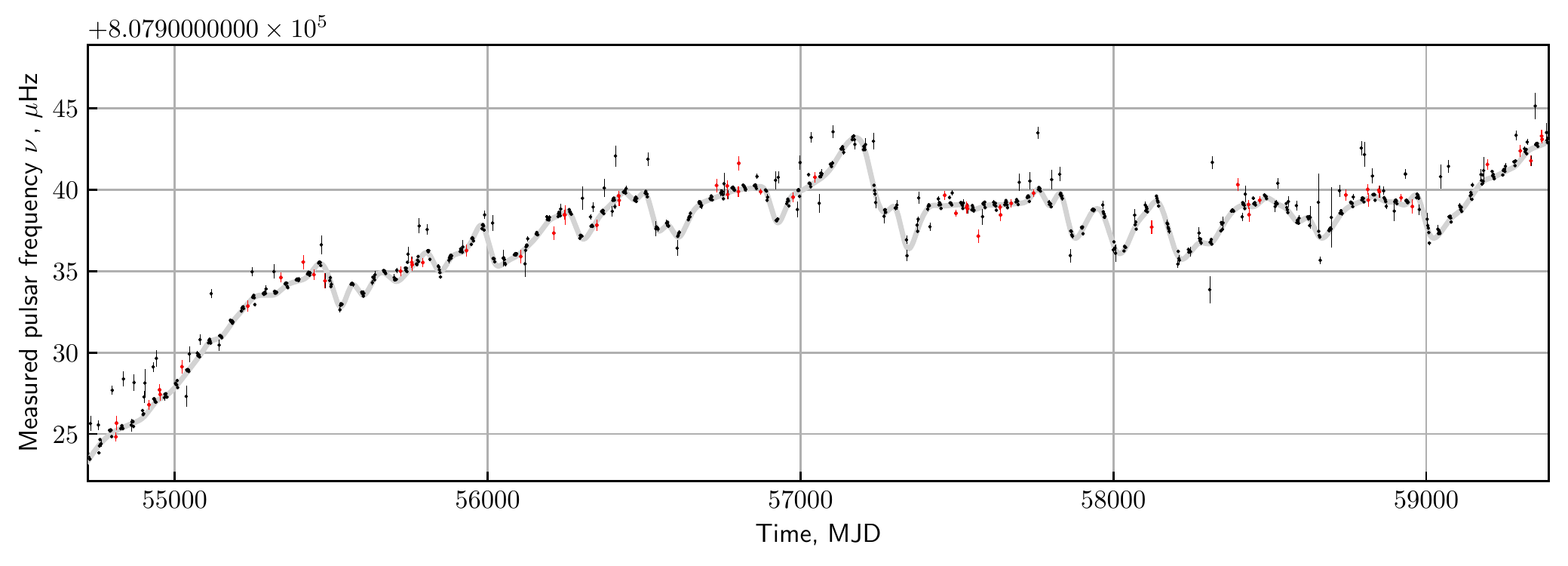}
    \caption{Her X-1 pulse frequency as a function of time. The dots show \textit{Fermi}/GBM measurements (in black during the 35-day cycle Main-on phases 0.0--0.35, in red otherwise). The grey solid line shows the long-term pulse frequency variations $\nu_0(t)$ approximated as described in Section \ref{sec:modelling}. }
    \label{f:pulsar-frequency}
\end{figure*}

It is straightforward to calculate analytically the pulse frequency variations from a freely precessing axially symmetric NS when the NS moments of inertia $I_1 = I_2 \neq I_3$ \citep[see, e.g,][]{1970Natur.225..838R, 1986ApJ...300L..63T, 1989A&A...221L...7B, 1993A&A...267L..43B, 1995pns..book...55S}. Below we will assume that the precession frequency is much lower than the spin frequency of the NS so that the total angular momentum vector to a high accuracy coincides with the NS spin vector. 
When the NS spin frequency vector $\boldsymbol{\Omega}$ is misaligned with the principal inertia axis $I_3$ by angle $\gamma$, the free precession angular frequency reads
\begin{equation}
     \omega = \Omega\,\frac{I_1 - I_3}{I_1} \cos{\gamma}\,.
    \label{e:Omegap}
\end{equation}
The observed pulse frequency is modulated  by the time derivative of the angle $\Phi$ marking the NS precession phase (see Fig. \ref{f:neutron_star}). For the angle $\beta$ between the north magnetic pole $N$ and $I_3$ axis, the phase $\Phi$ can be found from the sine and cosine theorem for spherical triangles:
\begin{equation}
    \cos{\Phi(t)} = \frac{\sin{\beta}\sin{\varphi(t)}}{\sqrt{1 - [\cos{\gamma}\cos{\beta} + \sin{\gamma} \sin{\beta} \cos{\varphi(t)}]^2}}\,,
\end{equation}
where $\varphi(t)$ is the azimuthal angle of the vector $\boldsymbol{\Omega}$ in a rigid coordinate frame related to the NS's principal inertia axes (the light grey lines in Fig. \ref{f:neutron_star}). In the course of NS free precession, $\varphi(t)$ is a linear function of time: 
\begin{equation}
    \varphi(t) = \varphi_0 + \Omega t\,.
    \label{eq:phi_t}
\end{equation}

The amplitude of the periodic sub-microsecond pulse frequency periodic variations observed by \textit{Fermi}/GBM in Her X-1 can be easily adjusted
by assuming a two-axial NS free precession with the appropriate choice of the NS ellipticity $\Delta I/I$ \citep{2021ARep...65.1039S}. However, the shape of the measured pulse frequency variations as a function of the 35-day phase can be better reproduced by assuming a slight NS triaxiality, $I_1\ne I_2\ne I_3$.


\subsection{Precession of a triaxial NS}
\label{sec:periodical-pulsar-frequency-variations}

Given the moments of inertia $I_1 < I_2 < I_3$ and angular velocity $\boldsymbol{\Omega}$, the NS rotational energy is 
\begin{equation}
    2E = I_1\Omega_1^2 + I_2\Omega_2^2 + I_3\Omega_3^2,
\end{equation}
and the angular momentum is
\begin{equation}
    M^2 = I_1^2\Omega_1^2 + I_2^2\Omega_2^2 + I_3^2\Omega_3^2.
\end{equation}
Following \cite{LANDAU197696}, the motion of the angular momentum vector is described by the equations
\begin{align}
    \Omega_1 &= \sqrt{\frac{2EI_3-M^2}{I_1(I_3-I_1)}}\,\mathrm{cn}\tau\\
    \Omega_2 &= \sqrt{\frac{2EI_3-M^2}{I_2(I_3-I_2)}}\,\mathrm{sn}\tau\\
    \Omega_3 &= \sqrt{\frac{M^2-2EI_1}{I_3(I_3-I_1)}}\,\mathrm{dn}\tau,
\end{align}    
where $\mathrm{cn}\tau$, $\mathrm{sn}\tau$, $\mathrm{dn}\tau$ are elliptic Jacobi functions,  and the dimensionless time $\tau$ is
\begin{equation}
    \tau = t\,\sqrt{\frac{(I_3-I_2)(M^2-2EI_1)}{I_1 I_2 I_3}}.
\end{equation}
The free precession period reads
\begin{equation}
\label{e:T}
    P = 4 \sqrt{\frac{I_1 I_2 I_3}{(I_3-I_2)(M^2-2EI_1)}} \int_{0}^{\pi/2} \frac{du}{\sqrt{1-k^2 \sin^2{u}}}\,,
\end{equation}
where the parameter $k$ is defined as
\begin{equation}
    k^2 = \frac{(I_2-I_1)(2EI_3-M^2)}{(I_3-I_2)(M^2-2EI_1)}\,.
\end{equation}

For a given NS rotational period, the fractional moment inertia differences $\Delta I_2 = (I_2-I_1)/I_1$ and $\Delta I_3 = (I_3 - I_1)/I_1$ fully determine the NS free precession period $P$. However, a realistic NS is not a fully rigid body. In Her X-1, the NS free precession period can change due to the action of external torques, mass accretion, non-rigid coupling between the crust and the core, etc.

\begin{figure*}
    \centering
    \includegraphics[width=\hsize]{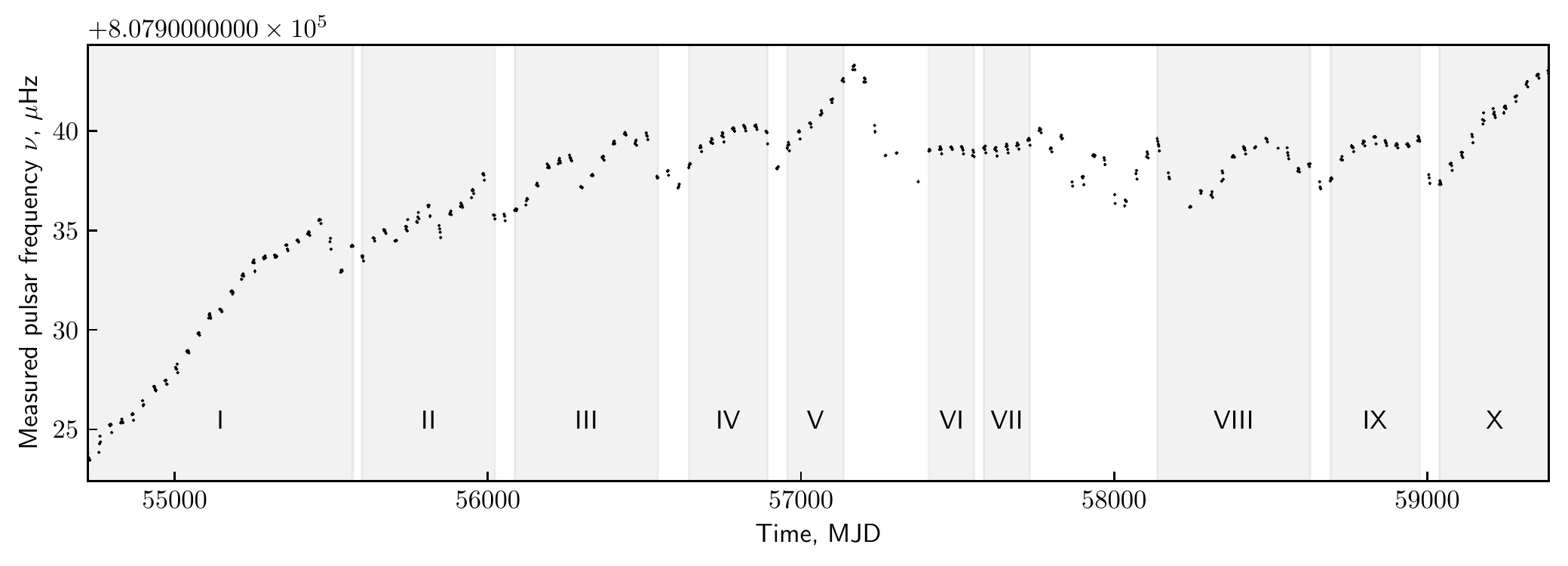}
    \caption{The same as in Fig. \ref{f:pulsar-frequency} with grey zones showing intervals $\Delta T_k$ with constant NS free precession period $P_k$  (see Table \ref{tab:parameters}). Also, measurements with error larger than $0.1 \mu\textrm{Hz}$ or outside Main-on state were removed.}
    \label{f:pulsar-frequency-zones}
\end{figure*}

%
In the \textit{Fermi}/GBM data, we identified 10 time intervals $\Delta T_k$, $k=\textrm{I},\ldots,\textrm{X}$ comprising $\sim 5-20$ consecutive cycles that can be described by approximately constant $P_k$ (see below Fig. \ref{f:pulsar-frequency-zones} and Table \ref{tab:parameters}). (Due to scarce points in some cycles, the data chunks $\Delta T_k$ with constant 35-day cycle duration are not always contingent and can be separated by time intervals, which we excluded from the analysis; their inclusion does not change the results but worsens the $\chi^2$ of the fit). We assigned equal values of $\Delta I_2$ for all 35-day cycles to minimize residuals between the model and observations. The parameter $(\Delta I_3)_k$ was calculated individually from Eq. (\ref{e:T}) inside each data intervals $\Delta T_k$ with constant period $P_k$. Thus, inside each data interval, for given NS parameters and free precession period $P_k$ we can numerically calculate positions of the vector $\boldsymbol{\Omega}$, the phase angle $\Phi$ (see Fig. \ref{f:neutron_star}) and derivative $d\Phi/dt$ defining the pulse frequency variations.



\section{Modelling of Her X-1 pulsar frequency variations}\label{sec:modelling}

In accreting X-ray pulsars, the long-term pulse frequency variations $\nu_{0}(t)$ are caused by various factors,  
e.g. by variable accretion torque which are difficult to predict. Here, in order to subtract the long-term pulse frequency variations, we model $\nu_0(t)$ as a cubic spline passing through nodes $\tau_j\,, \nu_0(\tau_j)$ as follows. 


We introduce the residuals $R$ between the observed pulsar frequency measurements $\nu_i$ at moments $t_i$ and the theoretical model $\nu(t)$:
\begin{equation}
\label{e:R}
    R = \sum_{i} \left(\nu(t_i) - \nu_i\right)^2 \,.
\end{equation}
where the index $i$ runs through all frequency measurements, index $j$ corresponds to the 35-day cycles considered, see Table \ref{t:long-term-evolution} in Appendix A.
Our theoretical model $\nu(t)$ is the sum of the periodic 35-day pulsar frequency variations $d\Phi/dt$ due to the NS free precession and long-term trend $\nu_0(t)$: 
\begin{equation}
    \nu(t) = \frac{1}{2\pi} \frac{d\Phi(t)}{dt} + \nu_0(t)
\end{equation}

The time coordinate $\tau_j$ of the spline nodes is defined as the mean time of the pulse frequency measurement 
within the $j$-th Main-on:
\begin{equation}
    \tau_j = \frac{1}{N_j} \sum_i t_i \,,
\end{equation}
Here, $N_j$ is the number of 
observations within the $j$-th Main-on.
The spline value $\nu_0(\tau_j)$ is the difference between the mean pulse frequency and the model NS free precession frequency at the moment $\tau_j$:
\begin{equation}
    \nu_0(\tau_j) = \frac{1}{N_j} \sum_i \nu_i - \frac{1}{2\pi}\frac{d\Phi(\tau_j)}{dt}\,,
\end{equation}

Parameters of the long-term evolution $\nu_0(t)$ and 35-day variations of X-ray pulse frequency $\delta\nu(t)$ were evaluated by minimizing the residuals $R$, equation \ref{e:R}. Parameters of the triaxial NS free precession are listed in Tables \ref{t:3-free-precession-parameters} and \ref{tab:parameters}.
The minimizing of the residuals $R$ were done using the LMFIT package \citep{newville_matthew_2014_11813}.

Inside each $k$-th data interval with constant 35-day cycle duration $P_k$, the fractional NS moment of inertia difference $\Delta I_3$ was optimized to fit the observed pulse frequency variations measured by \textit{Fermi}/GBM. 
The parameters $\Delta I_2$ and the NS principal axis of inertia $I_3$ misalignment with the angular momentum $\gamma_0$ were fixed for all 35-day cycles.


The trajectory of the NS angular momentum $\mathbf{\Omega}$ on the surface (see Fig. \ref{f:neutron_star}) can be defined by $\Delta I_2, \Delta I_3$ and the misalignment angle $\gamma_0$ at the NS precession zero phase (cf. Eq. (\ref{e:Omegap}) for two-axial case, where this angle is constant). With fixed $\Delta I_2$ and $\gamma_0$, the NS free precession period $P_k$ (Eq. \ref{e:T}) is defined by $\Delta I_3$ only. 
As seen from Table \ref{tab:parameters}, the 35-day period in Her X-1 changes within the range $34^d.8-35^d.2$, i.e. $|\Delta P/P|\simeq 1\%$ on a timescale of half a year or longer.
Variations of the moments of inertia are possible for a not fully rigid NS body; variations of the misalignment between the NS principal inertia axis $I_3$ and angular momentum can be due to the internal coupling between the NS crust and core. Both cases are physically plausible for a realistic NS. In our model with fixed $\gamma_0$, the changes in NS moments of inertia can be due to the redistribution of mass accreted onto the NS. Indeed, on a year timescale, the accreted mass in Her X-1 is $\delta M\sim 10^{17}\mathrm{[g/s]}\times 3\times 10^7[\mathrm{s}]\sim 3\times 10^{24}$~g, i.e. the fractional change in the NS moment of inertia is $\delta I/I\approx \delta M/M\sim 10^{-9}$. Thus, the mass redistribution in the non-rigid NS body with a mean ellipticity of $10^{-6}$ could be sufficient to produce $\lesssim 1\%$ variations in the relative difference of the NS moments of inertia (see Table \ref{tab:parameters}).

\begin{figure}
    \centering
    \includegraphics[width=\columnwidth]{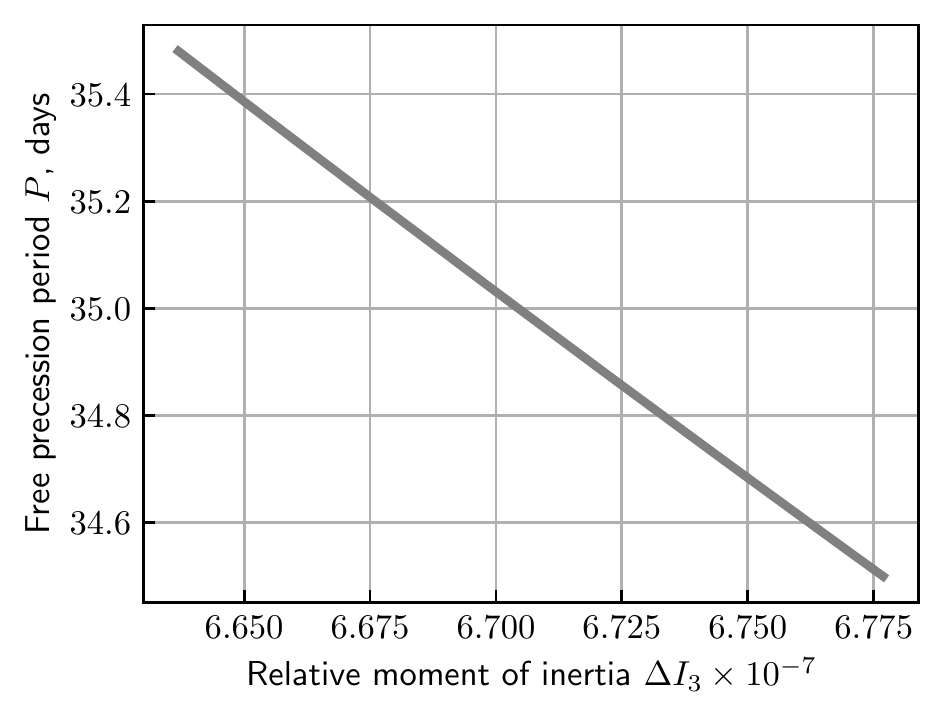}
    \caption{Triaxial NS free precession period $P$ as a function of relative moment of inertia difference $\Delta I_3$ with other parameters fixed, see Table \ref{t:3-free-precession-parameters}.}
    \label{f:P}
\end{figure}

The NS free precession period $P$ as a function of $\Delta I_3$ in our model for Her X-1 with parameters from Table \ref{t:3-free-precession-parameters} is shown in Fig. \ref{f:P}.
It is seen that a 1\% variations in $\Delta I_3$ alter the free precession period $P$ correspondingly. Therefore, the NS free precession model for Her X-1 suggests a 1\% change in the NS body parameters on a year timescale.
Similar indications have been obtained earlier from the analysis of O-C behaviour of the mean 35-day cycle duration \citep{2013MNRAS.435.1147P}. 

\begin{figure*}
    \centering
    \includegraphics[scale=0.7]{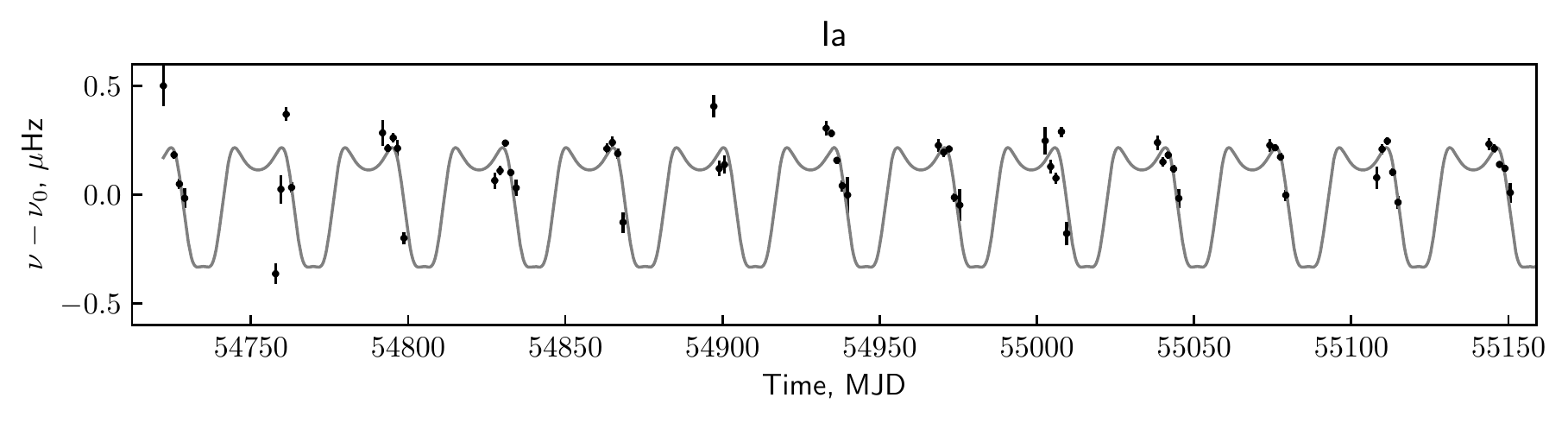}
    \includegraphics[scale=0.7]{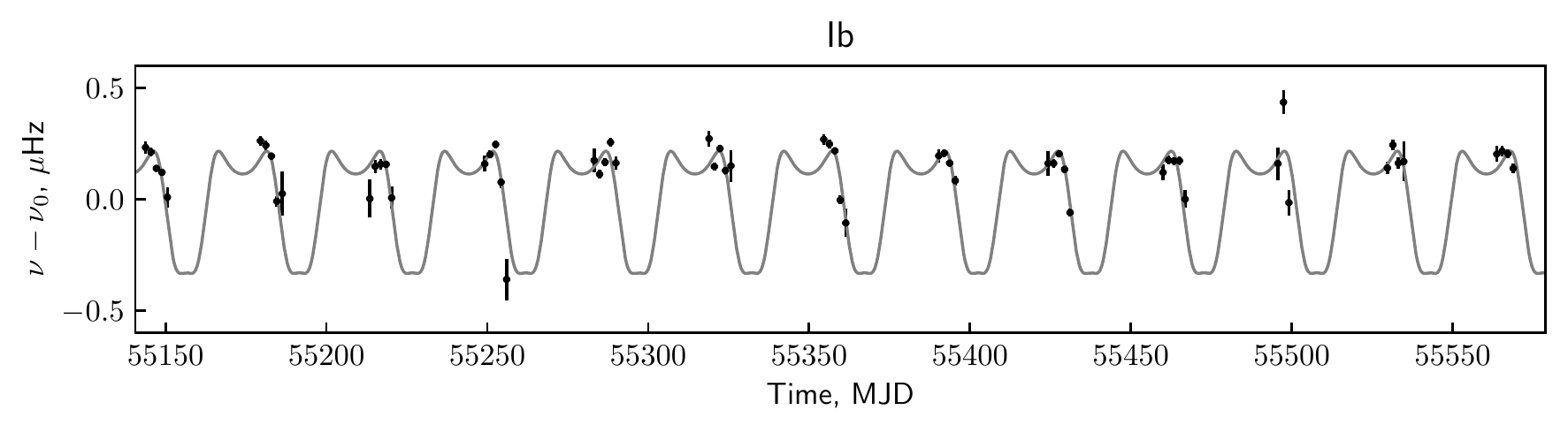}
    \includegraphics[scale=0.7]{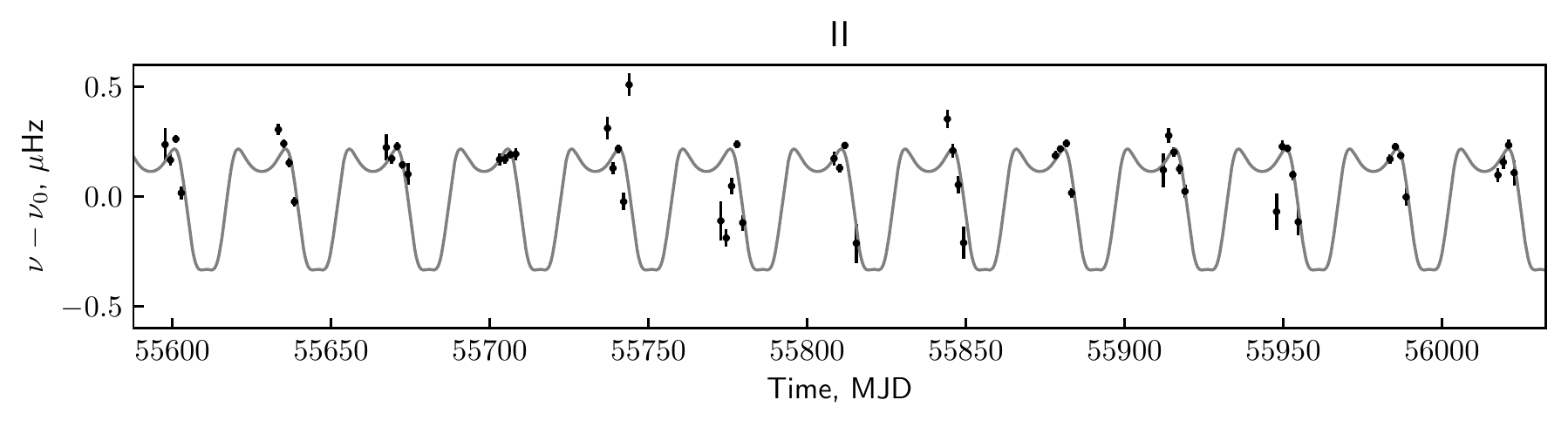}
    \includegraphics[scale=0.7]{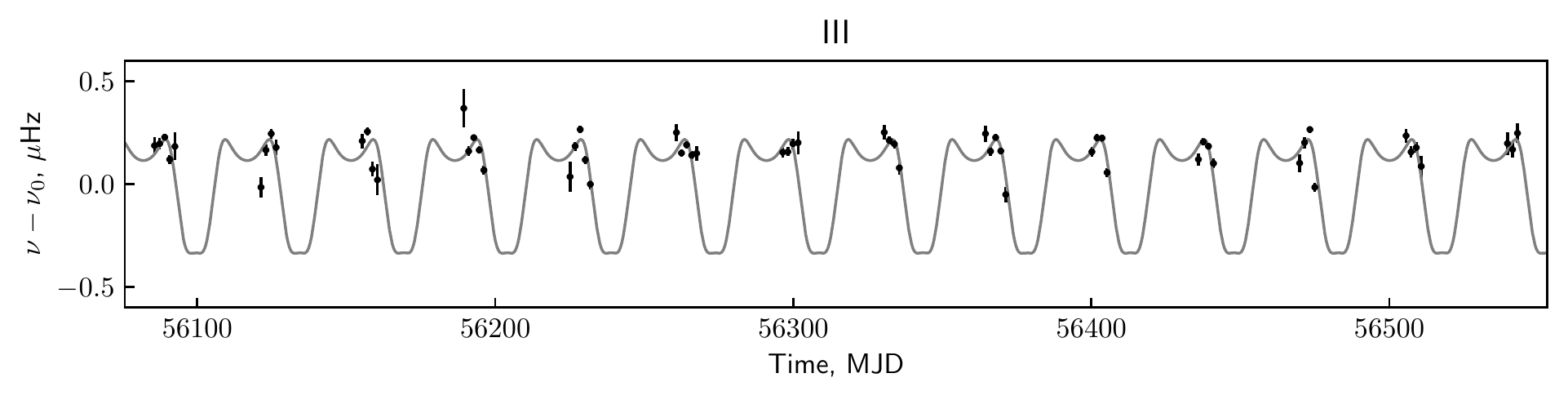}
    \includegraphics[scale=0.7]{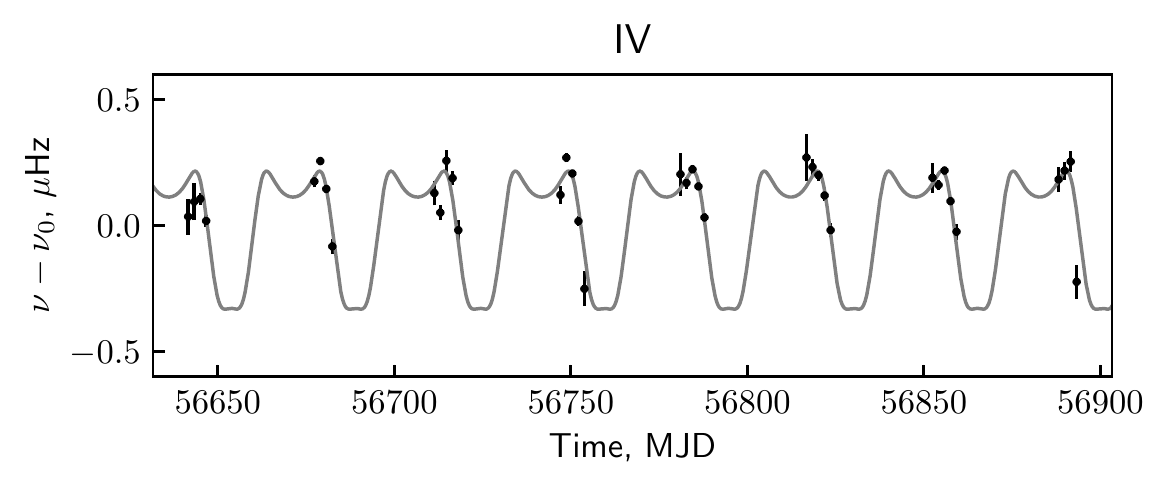}
    \includegraphics[scale=0.7]{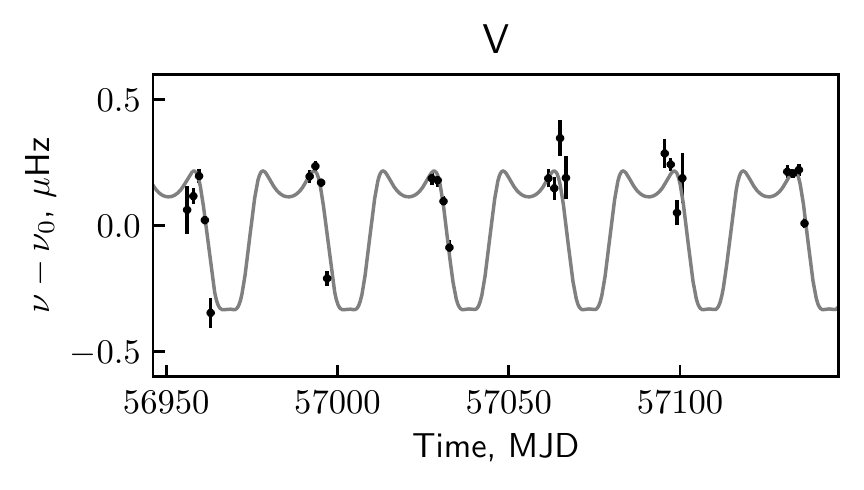}
    \includegraphics[scale=0.7]{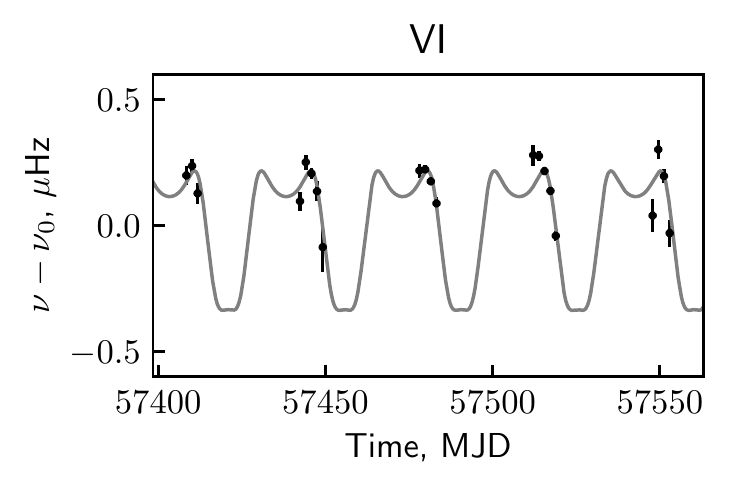}
    \includegraphics[scale=0.7]{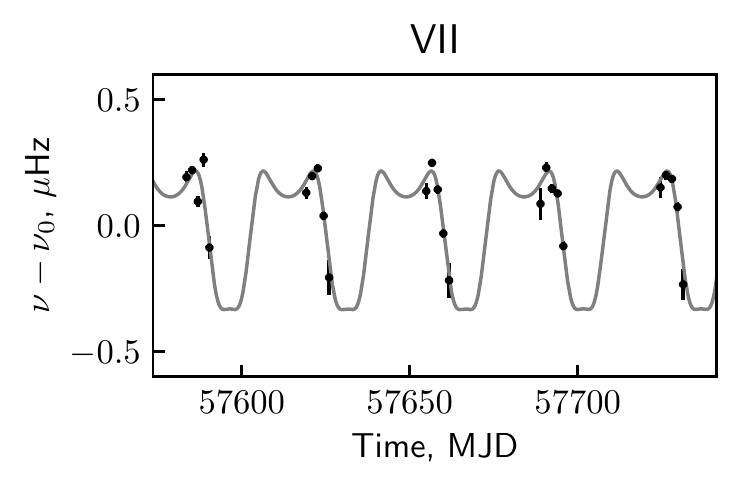}
    \caption{The best-fit modeling (the solid line) of the periodic X-ray pulse frequency variations of Her X-1 by the triaxial NS free precession (intervals I--VII from Table \ref{tab:parameters}).
   The dots show the \textit{Fermi}/GBM X-ray pulse frequency measurements.}
    \label{fig:pulse_variations-a}
\end{figure*}

\begin{figure*}
    \centering
    \includegraphics[scale=0.7]{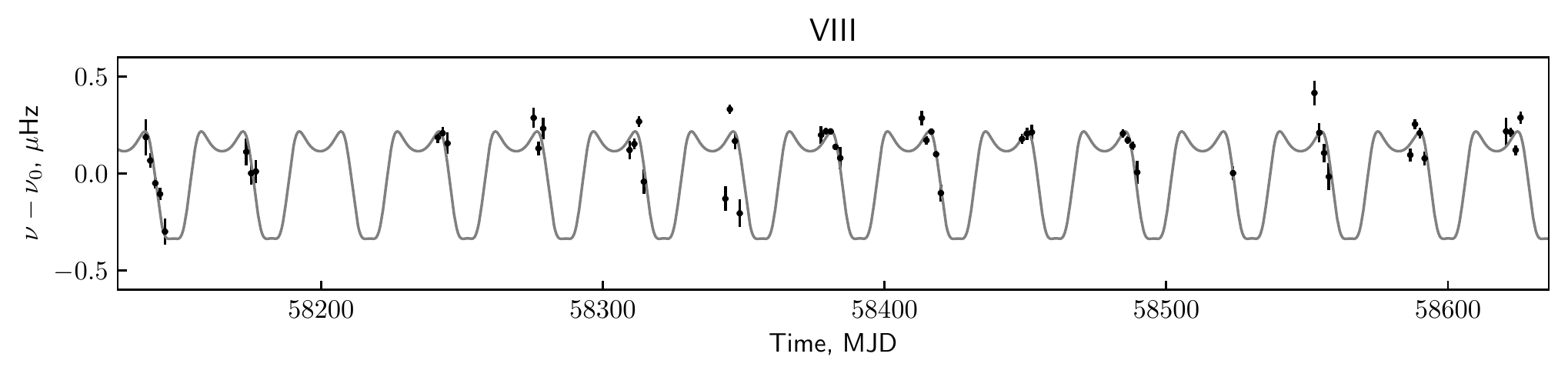}
    \includegraphics[scale=0.7]{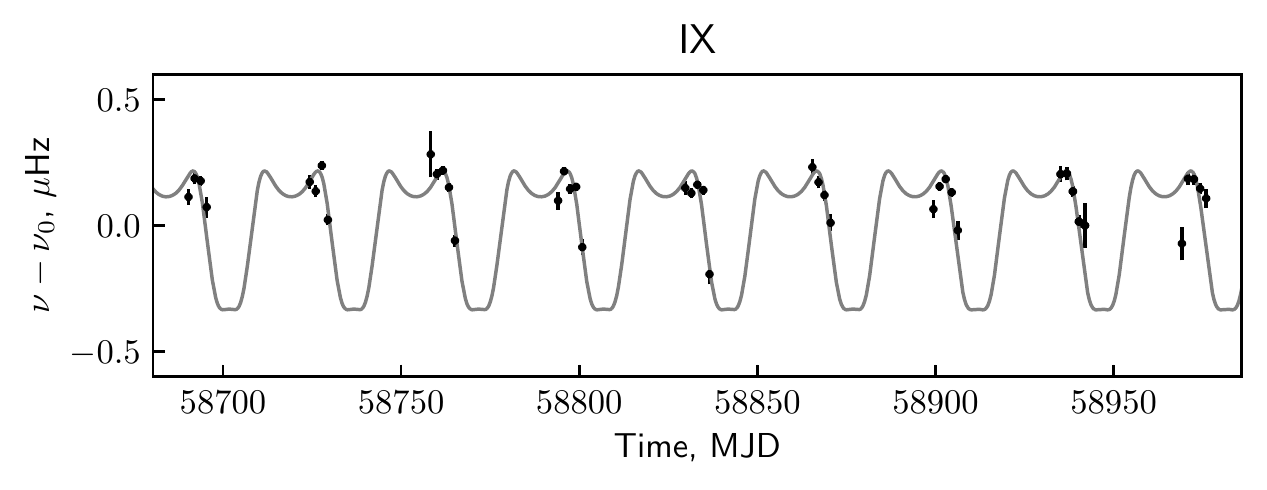}
    \includegraphics[scale=0.7]{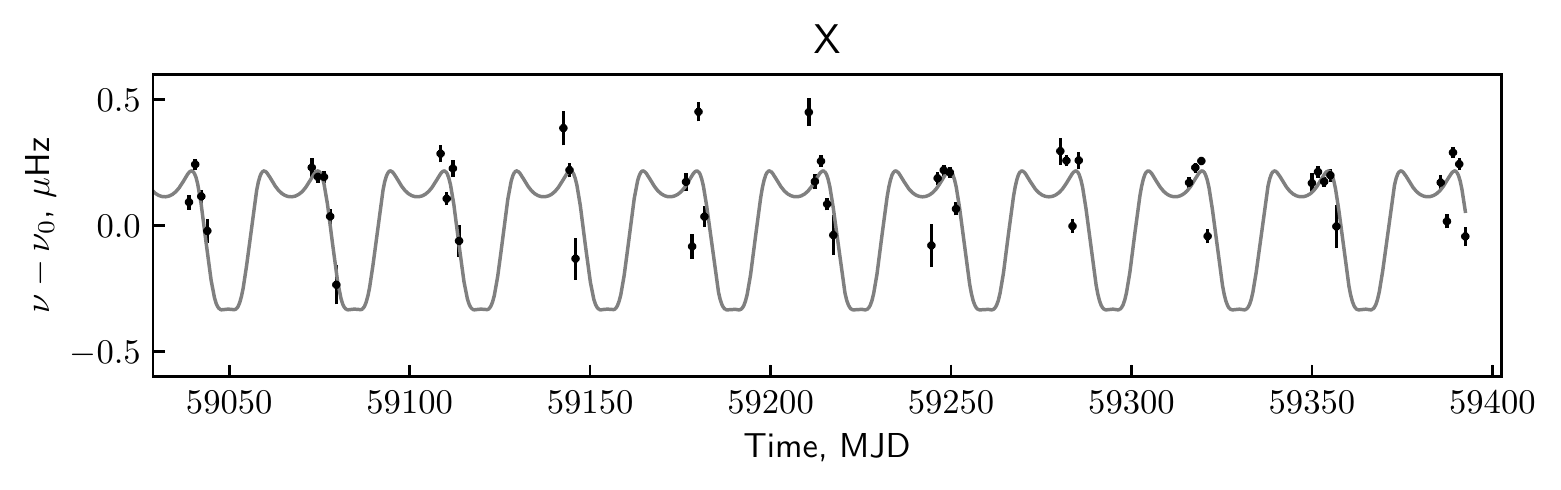}
    \caption{The same as in Fig. \ref{fig:pulse_variations-a} for
    intervals VIII--X from Table \ref{tab:parameters}.}
    \label{fig:pulse_variations-b}
\end{figure*}

The best-fit modeling of the periodic X-ray pulse variations of Her X-1 by the triaxial NS free precession 
with parameters from Table \ref{t:3-free-precession-parameters} and Table \ref{tab:parameters} is shown in Figs. \ref{fig:pulse_variations-a} and \ref{fig:pulse_variations-b}.  The solid  black line presents the model $\nu(t)$, with the 35-day cycle duration $P_k$ adjusted using the fractional moment inertia difference $\Delta I_3$ and the NS free precession zero phase at the beginning of each data interval $\Delta T_k$ listed in Table  \ref{tab:parameters}.

\begin{table}
  \caption{Triaxial free precession model parameters fixed during the fitting inside $k$-th data intervals with constant 35-day cycle duration $P_k$}
  \label{t:3-free-precession-parameters}
  \begin{tabular}{l|l|l}
    \hline
    Parameter & Symbol & Value \\
    \hline
    $\boldsymbol{\Omega}$ and $I_3$ axis misalignment & $\gamma_0$ &  $50\degree$\\
    at zero free precession phase                   & &  \\
    \hline
    Coordinates of the  & $N_{\phi}$  &   $90\degree$\\
    magnetic pole $N$                                          & $N_{\theta}$ &  $30\degree$\\
\hline
Fractional moment inertia difference $(I_2 - I_1)/I_1$ & $\Delta I_2$ & $3\times 10^{-7}$ \\
\hline
  \end{tabular}

\end{table}

\begin{table*}
    \centering
    \caption{Triaxial NS free precession model parameters inside $k$-th data intervals with constant 35-day cycle duration $P_k$ marked in Fig. \ref{f:pulsar-frequency-zones}.}
    \begin{tabular}{l|c|c|c|c|c}
        \hline
         Interval number $k$ & $\Delta T_k$, MJD & Cycle duration, $P_k$ & $\Delta I_3 \times 10^{-7}$ & $\varphi_0^{*}$ & Reduced $\chi^2$  \\
         \hline
         I    & 54722.15 -- 55568.84 & $35.14$ &  6.68 & $-0.45$ & $7.2$ \\
         II   & 55597.73 -- 56022.78 & $35.02$ &  6.70 & $-0.305$ & $9.3$ \\
         III  & 56085.66 -- 56543.01 & $34.85$ &  6.73 & $0.024$ & $4.9$ \\
         IV   & 56641.62 -- 56893.27 & $35.25$ &  6.67 & $-1.13$ & $3.2$ \\
         V    & 56956.11 -- 57136.34 & $35.05$ &  6.70 & $-0.63$ & $4.4$ \\
         VI   & 57408.42 -- 57552.93 & $34.83$ &  6.73 & $0.01$ & $1.8$\\
         VII  & 57583.53 -- 57731.39 & $35.1$  &  6.69 & $-0.88$ & $4.8$ \\
         VIII & 58137.80 -- 58625.74 & $34.8$  &  6.73 & $-0.01 $ & $4.4$ \\
         IX   & 58690.33 -- 58975.96 & $35.01$ &  6.70 & $0.0$ & $3.6$ \\
         X    & 59039.85 -- 59392.50 & $35.0$  &  6.70 & $-0.015$ & $10.0$ \\
         \hline
        \multicolumn{6}{l}{$^{*}$ initial phase $\varphi_0$ (see Eq. \ref{eq:phi_t}) is calculated for the time of first \textit{Fermi}/GBM data point for Her X-1 $t_0 = $ MJD 54722.15470}
    \end{tabular}
    \label{tab:parameters}
\end{table*}

\section{Discussion \& Conclusion}

The 0.3--0.5 microsecond variability of the X-ray pulse period of Her X-1 measured by \textit{Fermi}/GBM\footnote{\url{https://gammaray.nsstc.nasa.gov/gbm/science/pulsars/lightcurves/herx1.html}} with $\sim 0.1$ microsec accuracy suggests the emitting region radial velocity amplitude $V_r/c \leq \Delta P^*/P^* = \Delta \nu^{*} / \nu^{*} \sim 3\times 10^{-7}$. In the present paper, we have shown that such variations are possible for a freely precessing, likely triaxial NS in Her X-1. In the \textit{Fermi}/GBM data, we have identified several time intervals with a duration of half a year or longer (see Fig. \ref{f:pulsar-frequency-zones} and Table \ref{tab:parameters}) where the NS precession period does not change noticeably. The NS precession period varies within 1\% in different intervals. Such variations can be explained by $\lesssim 1\%$ changes in the NS moment inertia difference due to accreted mass readjustment or variable internal coupling of the  NS crust with the core.

In principle, besides the NS free precession, the pulsar frequency variations could be generated by the reflection from the warped accretion disk precessing with the angular velocity $\Omega_{d}$. In that case, the maximum radial velocity of the reflector should be $V_{r,max}=\Omega_{d}R_{in}\approx 2\times 10^2$ cm/s for the assumed inner disk radius $R_{in}\approx 10^8$
cm. This velocity would give rise to the Doppler frequency modulation with an amplitude of $\Delta \nu_{ref}/\nu^*\sim 10^{-8}$, much smaller than the observed value.
The Doppler broadening of the reprocessed pulsations on the accretion disk flow would smear the precession-induced frequency variations. 
There is another point of concern with the disk reflection model. In Her X-1, the beginning of the 35-day cycle is known to be due to the central X-ray source opening by the outer parts of the precessing accretion  disk \citep{2005A&A...443..753K}. If the pulse period variations were produced by the reflection from the disk, one would expect correlation between the 35-day cycle beginning and the pulse frequency maximum, which is not found.
Therefore, the possibility that the observed pulsar period change in Her X-1 is due to reprocession of the X-ray pulses on the disk seems unlikely.

In our model, the inner part of the disk should align with the NS's equator due to magnetic forces \citep{1980SvAL....6...14L,  1981AZh....58..765L, 1999ApJ...524.1030L} during the 35-day cycle Main-on. The pulsar period $1.24$ s should be close to the equilibrium value (the magnetospheric radius is close to the corotation radius), suggesting the inner disk radius $\sim 100 \,R_{\textrm{ns}}$. Therefore, the accreting plasma gets frozen into the magnetic field and is canalised onto the NS's surface in regions defined by the local magnetic field structure. In this case, the precession of the outer parts of the disk should not produce variations of the hot spot geometry.

During the Short-on stage, the X-ray flux from Her X-1 is several times as low as at the Main-on, and the pulse period determination from \textit{Fermi}/GBM data is less certain. However, on several occasions (e.g., on MJD 54952, 55757, 56418, 57532) the pulse period is found to be at the approximately the same level as at the Main-on$^2$. In our model, the Short-on pulse is shaped by emitting arcs located symmetrically to the inertia axis $I_3$ but phase-separated by $\pi$ \citep[see Fig. 2 and 3 in][]{2013MNRAS.435.1147P}. Therefore, the expected pattern of the pulse profile variations during the Short-on should be similar to the Main-on. Future accurate measurements of the X-ray pulse timing in Her X-1 Short-on are valuable to test this prediction.

We conclude that a freely precessing NS in Her X-1 with parameters inferred from an independent  analysis of X-ray pulse profile evolution with 35-day phase \citep{2013MNRAS.435.1147P} can explain regular sub-microsecond pulse period changes observed by \textit{Fermi}/GBM. 
To explain a $\lesssim 1\%$ variations in the NS free precession period on a year timescale, the model requires the corresponding change in the NS parameters (relative difference in the moments of inertia or the NS angular momentum misalignment with the principal moment of inertia). These changes might be related to the variable internal coupling of the NS crust with the core.
The model has also proved successful in explaining the HZ Her optical light curves over the 35-day cycle as well \citep{2020MNRAS.499.1747K}. Therefore, after about half century of studies, the NS free precession as the inner clock mechanism for the observed 35-day cycle in Her X-1/HZ Her is further supported by the X-ray pulse period frequency variations observed by \textit{Fermi}/GBM.

\section*{Acknowledgements}
We thank the anonymous referee for useful comments.
The work of DK and NS was supported by the RSF grant 21-12-00141 (modelling of Her X-1 pulsar frequency variations; calculation of \textit{Swift}/BAT 35-day cycle turn-on times). 
The authors acknowledge the Interdisciplinary Scientific Educational School of Moscow University 'Fundamental and applied space research'. KP acknowledges support by the Kazan Federal University Strategic Academic Leadership Program ("PRIORITY-2030").

\section*{Data Availability}
The data underlying this article are available in the article, \textit{Fermi}/GBM X-ray data are freely available at \url{https://gammaray.nsstc.nasa.gov/gbm/science/pulsars/lightcurves/herx1.html}, \textit{Swift}/BAT X-ray data are freely available at \url{https://swift.gsfc.nasa.gov/results/transients/HerX-1/}.

\bibliographystyle{mnras}
\bibliography{bibliography}

\begin{appendix}

\section{Her X-1 Long-term pulse frequency evolution}\label{sec:app-long-term-evolution}

Here we present the table of the spline values $\tau_j\,, \nu_0(\tau_j)$ of the long-term evolution of Her X-1 pulse frequency. The method of calculation of $\tau_j\,,\nu_0(\tau_j)$ is described in Section \ref{sec:modelling}. The numbering of 35-day cycles follows the convention introduced by \cite{1983A&A...117..215S}.


\begin{table}
\caption{Pulse frequency long-term evolution}
\label{t:long-term-evolution}
\begin{tabular}{lcc}
\hline
Cycle number $j$  & $\tau_j$ MJD & $\nu_0(\tau_j) \times 10^{-5} + 0.8079$, Hz \\
\hline
 383  &    54726.91 &  2.3476 \\ 
 385  &    54760.36 &  2.4355 \\ 
 386  &    54796.05 &  2.5070 \\ 
 387  &    54831.80 &  2.5380 \\
 388  &    54865.80 &  2.5659 \\
 389  &    54900.66 &  2.6082 \\
 390  &    54936.37 &  2.6978 \\
 391  &    54971.21 &  2.7340 \\
 392  &   55006.07 &  2.8031 \\
 393  &   55041.76 &  2.8834 \\
 394  &   55076.63 &  2.9694 \\
 395  &   55112.33 &  3.0662 \\
 396  &   55146.33 &  3.0910 \\
 397  &   55182.04 &  3.1806 \\
 398  &   55216.88 &  3.2679 \\
 399  &   55251.74 &  3.3362 \\
 400  &   55287.45 &  3.3517 \\
 401  &   55321.46 &  3.3598 \\
 402  &   55357.15 &  3.4129 \\
 403  &   55392.85 &  3.4379 \\
 404  &   55428.56 &  3.4796 \\
 405  &   55464.27 &  3.5462 \\
 406  &   55498.27 &  3.4341 \\
 407  &   55531.43 &  3.2819 \\
 408  &   55567.13 &  3.4153 \\
 409  &   55601.13 &  3.3524 \\
 410  &   55635.97 &  3.4482 \\
\hline
\end{tabular}
\end{table}

\begin{table}
\contcaption{Pulse frequency long-term evolution}
\begin{tabular}{lcc}
\hline
Cycle number $j$  & $\tau_j$ MJD & $\nu_0(\tau_j) \times 10^{-5} + 0.8079$, Hz \\
\hline
 411  &   55670.85 &  3.4878 \\
 412  &   55705.68 &  3.4367 \\
 413  &   55741.39 &  3.4997 \\
 414  &   55777.10 &  3.5698 \\
 415  &   55810.62 &  3.6085 \\
 416  &   55845.94 &  3.4974 \\
 417  &   55880.79 &  3.5782 \\
 418  &   55916.50 &  3.6168 \\
 419  &   55951.36 &  3.6936 \\
 420  &   55986.21 &  3.7729 \\
 421  &   56019.37 &  3.5627 \\
 422  &   56053.36 &  3.5510 \\
 422  &   56089.08 &  3.5909 \\
 424  &   56123.93 &  3.6445 \\
 425  &   56157.93 &  3.7143 \\
 426  &   56193.66 &  3.8105 \\
 427  &   56229.34 &  3.8463 \\
 428  &   56264.19 &  3.8520 \\
 429  &   56299.04 &  3.7043 \\
 430  &   56333.04 &  3.7663 \\
 431  &   56368.75 &  3.8619 \\
 432  &   56403.61 &  3.9323 \\
 433  &   56438.46 &  3.9829 \\
 434  &   56473.31 &  3.9383 \\
 435  &   56508.17 &  3.9673 \\
 436  &   56541.32 &  3.7576 \\
 437  &   56575.32 &  3.7831 \\
 438  &   56609.32 &  3.7088 \\
 439  &   56645.02 &  3.8269 \\
 440  &   56679.87 &  3.9120 \\
 441  &   56714.75 &  3.9438 \\
 442  &   56750.44 &  3.9712 \\
 443  &   56785.31 &  3.9958 \\
 444  &   56821.00 &  4.0038 \\
 445  &   56856.69 &  4.0148 \\
 446  &   56891.56 &  3.9804 \\
 447  &   56924.70 &  3.7997 \\
 448  &   56959.57 &  3.9324 \\
 449  &   56994.40 &  3.9863 \\
 450  &   57030.10 &  4.0276 \\
 451  &   57064.13 &  4.0673 \\
 452  &   57099.84 &  4.1381 \\
 453  &   57133.83 &  4.2445 \\
 454  &   57169.52 &  4.3165 \\
 455  &   57204.38 &  4.2516 \\
 456  &   57236.69 &  3.9947 \\
 457  &   57269.85 &  3.8602 \\
 458  &   57305.56 &  3.8793 \\
 459  &   57339.56 &  3.6466 \\
 460  &   57375.24 &  3.7617 \\
 461  &   57410.11 &  3.8909 \\
 462  &   57444.96 &  3.9009 \\
 463  &   57480.66 &  3.9054 \\
 464  &   57516.37 &  3.9028 \\
 465  &   57551.23 &  3.8788 \\
 466  &   57586.08 &  3.9012 \\
 467  &   57621.77 &  3.9046 \\
 468  &   57657.47 &  3.9111 \\
 469  &   57693.46 &  3.9164 \\
 470  &   57728.89 &  3.9432 \\
 471  &   57762.88 &  3.9992 \\
 472  &   57798.58 &  3.9025 \\
 473  &   57832.60 &  3.9695 \\
 \hline
\end{tabular}
\end{table}

\begin{table}
\contcaption{Pulse frequency long-term evolution}
\begin{tabular}{lcc}
\hline
Cycle number $j$  & $\tau_j$ MJD & $\nu_0(\tau_j) \times 10^{-5} + 0.8079$, Hz \\
\hline
 474  &   57865.75 &  3.7353 \\
 475  &   57900.60 &  3.7525 \\
 476  &   57935.46 &  3.8704 \\
 477  &   57969.45 &  3.8382 \\
 478  &   58003.45 &  3.6421 \\
 479  &   58036.61 &  3.6337 \\
 480  &   58070.63 &  3.7774 \\
 481  &   58105.48 &  3.8695 \\
 482  &  58141.17 &  3.9264 \\
 483  &  58175.18 &  3.7653 \\
 484  &  58208.36 &  3.5862 \\
 485  &  58243.18 &  3.6102 \\
 486  &  58277.18 &  3.6743 \\
 487  &  58312.04 &  3.6719 \\
 488  &  58346.89 &  3.7699 \\
 489  &  58380.89 &  3.8702 \\
 490  &  58416.61 &  3.9037 \\
 491  &  58450.59 &  3.8997 \\
 492  &  58486.31 &  3.9490 \\
 493  &  58521.13 &  3.9164 \\
 494  &  58556.02 &  3.8763 \\
 495  &  58589.17 &  3.8031 \\
 496  &  58624.88 &  3.8219 \\
 497  &  58658.09 &  3.7112 \\
 498  &  58692.78 &  3.7505 \\
 499  &  58726.88 &  3.8553 \\
 500  &  58762.58 &  3.9048 \\
 501  &  58797.45 &  3.9344 \\
 502  &  58832.30 &  3.9583 \\
 503  &  58868.00 &  3.9290 \\
 504  &  58902.00 &  3.9213 \\
 505  &  58937.69 &  3.9152 \\
 506  &  58973.42 &  3.9562 \\
 507  &  59006.56 &  3.7438 \\
 508  &  59041.41 &  3.7264 \\
 509  &  59075.41 &  3.8179 \\
 510  &  59110.27 &  3.8715 \\
 511  &  59144.27 &  3.9584 \\
 512  &  59179.14 &  4.0506 \\
 513  &  59213.98 &  4.0771 \\
 514  &  59248.84 &  4.1086 \\
 515  &  59283.68 &  4.1487 \\
 516  &  59318.54 &  4.2293 \\
 517  &  59352.54 &  4.2662 \\
 518  &  59389.16 &  4.3000 \\
 \hline
\end{tabular}
\end{table}

\end{appendix}
\label{lastpage}
\end{document}